\documentstyle[aps,prb,preprint,epsf]{revtex}
\title{Non-interacting Cooper pairs inside a pseudogap.}
\author{
Oleg Tchernyshyov\footnote{e-mail address: olegt@cuphyb.phys.columbia.edu}
}
\address{Department of Physics, Columbia University, New York, NY 10027}
\date{To appear in Physical Review B}
\begin{document}
\maketitle

\begin{abstract}
I present a simple analytical model describing the normal state of 
a superconductor with a pseudogap in the density of states, such as in 
underdoped cuprates.  
In nearly two-dimensional systems, where the superconducting transition
temperature is reduced from the mean-field BCS value, 
Cooper pairs may be present as slow
fluctuations of the BCS pairing field.  Using the self-consistent 
$T$-matrix (fluctuation exchange) approach I find that 
the fermion spectral weight exhibits two BCS-like peaks, broadened
by fluctuations of the pairing field amplitude.  
The density of states becomes suppressed near 
the Fermi energy, which allows for long-lived low-energy Cooper pairs that
propagate as a sound-like mode with a mass.  A self-consistency requirement,
linking the width of the pseudogap to the intensity of the pairing field,
determines the pair condensation temperature.  In nearly
two-dimensional systems, it is proportional to the degeneracy temperature
of the fermions, with a small prefactor that vanishes in two dimensions.  
\end{abstract}


\section{Introduction}

One of the strange features of high-temperature cuprate superconductors,
a normal-state gap in the density of states, remains a subject of 
controversy.  A great deal of experimental evidence (NMR,\cite{NMR} 
optical conductivity,\cite{IR} specific heat,\cite{Loram} and, 
most recently, 
angle-resolved photoemission\cite{Loeser,Ding}) seems to indicate that 
the superconducting energy gap survives the transition to the normal state
and disappears only at a considerably higher temperature.  Even the 
anisotropic character of the gap 
is preserved.  This behavior is characteristic of ``underdoped'' 
compounds, in which doping of additional hole carriers increases 
the temperature of the superconducting transition $T_c$. 
There is no generally accepted theoretical model of this phenomenon,
although arguably the most popular tentative explanation puts the blame
on superconducting fluctuations near the transition point, which are 
expected to be enhanced in these highly anisotropic, almost two-dimensional,
materials.  While Monte-Carlo simulations of the 
attractive Hubbard model in two dimensions provide numerical 
evidence\cite{Randeria92} for the pairs-above-$T_c$ scenario, 
no analytical description of the pseudogap regime has been offered so 
far.  The goal of this paper is to provide such a sketch.  

Suppose that the pseudogap in cuprates does have the same origin 
as the superconducting gap --- the scattering between
fermion and hole states with the charge $\pm 2$ released in the form of a 
low-energy Cooper pair.  This picture rests on certain assumptions. 
Simply saying that Cooper pairs are present above $T_c$ as a propagating mode
is not enough.  It is important that the fluctuating pairing field 
$\Delta$ look frozen on the time scale $1/|\Delta|$, 
otherwise the pseudogap will not be formed.  Since the typical 
frequency of free propagating bosons in thermal equilibrium is given 
by their temperature, the pseudogap is expected to exist at temperatures
$T\ll T_0$, where $T_0$ is loosely defined to be of order $|\Delta|/k_B$.
In general, it may also be necessary to require the {\em spatial} coherence 
of the pairing field over a pair size $\xi_0\approx v/|\Delta|$.  In this 
particular model, however, the boson velocity is of order of the Fermi 
velocity $v$, so that as long as the pairing fluctuations are slow in 
time, they are also slow in space.  

In a somewhat arbitrary way, $T_0$ can be identified with
the transition temperature calculated in the BCS theory.  As in any other 
mean-field theory, the onset of a long-range order coincides with 
the appearance of a short-range order (formation of pairs), 
the assumption being that the phase of the order parameter will lock up
as soon as there is a non-zero amplitude.  This, of course,
breaks down in two spatial dimensions, as shown long ago by 
Hohenberg.\cite{Hohenberg}
In a highly anisotropic, almost two-dimensional system, 
long-range order sets in at a temperature $T_c\ll T_0$.  
Above $T_0$, Cooper pairs decay; between $T_0$ and $T_c$, they represent a 
propagating mode; finally, below $T_c$, they form a Bose condensate.  
In the intermediate range, at least when $T_c<T\ll T_0$, they represent a 
slowly fluctuating BCS pairing field and thus create a pseudogap.  
A well-defined pseudogap regime may be a peculiarity of low-dimensional
systems.  

A theoretical framework for treating the interplay between fermions and 
their boson-like bound states is thus needed.  Following pioneering works by 
Eagles,\cite{Eagles} Leggett,\cite{Leggett} and Nozieres and 
Schmitt-Rink,\cite{Nozieres} several approximate methods have been 
suggested: the boson-fermion model,\cite{FL,Ranninger} functional 
integration,\cite{Randeria,Pistolesi,Loktev} and the self-consistent
$T$-matrix approximation.\cite{Haussmann,Scalapino,Micnas}  In most cases, 
however, one has to resort to numerical computations.  The present work 
is written with the purpose to provide an analytical sketch of a
{\em normal state with a pseudogap}.  As such, it inevitably contains further 
simplifications, which hopefully do not alter the nature of the problem:
(1) The decay of low-energy bosons is neglected. 
(2) A clear separation of energy scales is assumed:
\begin{equation}
\label{separate scales}
k_B T_c \ll |\Delta| \ll \epsilon_F
\end{equation}
($\epsilon_F$ is the Fermi energy).  

Using the self-consistent $T$-matrix approach, I derive approximate 
propagators 
for fermions and Cooper pairs.  The width of the pseudogap is determined by 
the mean-square fluctuation of the pairing field 
$\langle|\Delta(x,x')|^2\rangle$ in a thermal ensemble.  This
expectation value depends, among other things, on the energy spectrum of 
Cooper pairs, which in turn is a function of the fermion energy 
spectrum.  A closed set of coupled equations results, allowing one to 
determine the condensation temperature of pairs.  
The so determined $T_c$ is proportional to the fermion density
and inverse mass, as seen on the Uemura\cite{Uemura plot} plot.
This is despite the fact that, as long as $|\Delta|\ll\epsilon_F$, the 
system is not in the limit of local (tightly bound) pairs.  
I also give a reason why quasiparticle peaks are quite broad 
near the Fermi surface, even when thermal fluctuations are slow 
($k_B T_c \ll |\Delta|$).  
If correct, this sketch may provide a simple way to understand two puzzling 
features of underdoped cuprate superconductors -- the pseudogap and the 
doping dependence of $T_c$ -- from a unified standpoint.  

The paper is organized as follows.  The conserving $T$-matrix 
approximation is outlined 
in Section \ref{T-matrix}; a bosonic propagator for a Cooper pair is 
introduced for the case of a ``separable'' interaction vertex.  
Section \ref{fermions} contains a derivation of the fermion propagator 
in the presence of a slowly fluctuating pairing field.  
An exact model of Section \ref{Gaussian fluctuations}
reveals an important difference between a true superconducting gap and 
a pseudogap, stemming from the quantum nature of pairing fluctuations. 
Low-energy properties of Cooper pairs in the presence of a pseudogap and 
their condensation temperature are derived in Section \ref{bosons}.
The effect of fast fluctuations is estimated in the Appendix.

\section{Self-consistent $T$-matrix approximation.}
\label{T-matrix}

A good description of the self-consistent $T$-matrix approximation can be 
found in Ref.~\ref{Haussmann}.  The relation to other approximate methods 
is outlined in Ref.~\ref{Micnas}.  

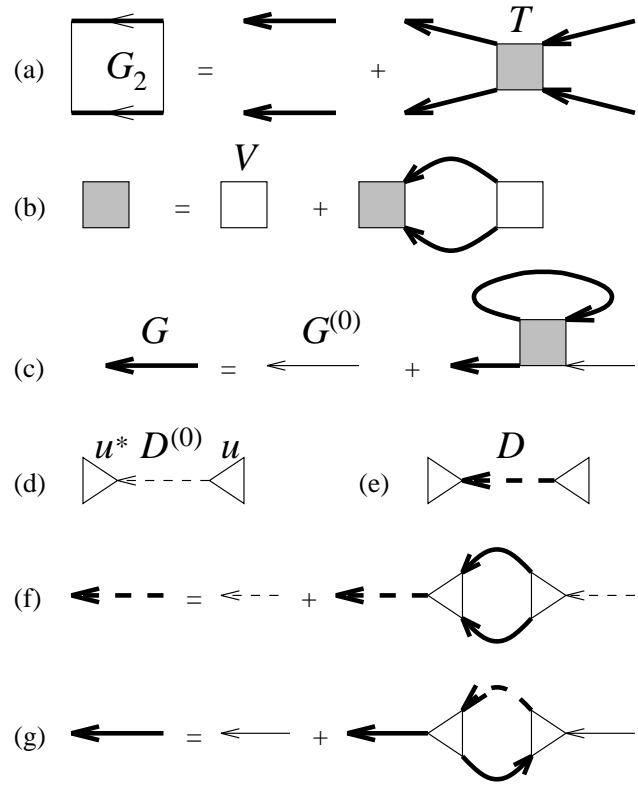
\begin{figure}
\caption{Diagrams of the self-consistent $T$-matrix approximation for the 
singlet channel.  
}
\label{graphs}
\end{figure}

The conserving $T$-matrix approximation is somewhat similar to the well-known 
Hartree-Fock principle.\cite{Baym}  The latter neglects any correlations
between interacting particles (except for statistical ones) and describes
the motion of independent entities in a self-consistent potential.  
The $T$-matrix approach includes pairwise correlations between
colliding particles, which are particularly important when two-particle 
bound states are formed.  The form of these correlations is inferred from the 
exact solution of a similar problem with two particles in vacuum, when the 
two-particle Green's function 
\begin{equation}
G_2(x,x';y',y)
\equiv-i\langle T[\psi(x)\psi(x')\psi^\dagger(y')\psi^\dagger(y)]\rangle
\end{equation}
can be expressed in terms of the one-particle
Green's function $G(x;y)\equiv -i\langle T[\psi(x)\psi^\dagger(y)]\rangle$
and the two-body $T$ matrix [Fig.~\ref{graphs}(a)].  In the many-body case, 
the latter is {\em defined} as a solution of the Bethe-Salpeter equation
[Fig.~\ref{graphs}(b)]:
\begin{equation}
\label{T}
T(P|k,k') = V(P|k,k') 
+ \frac{i}{\cal V}\sum_{k''}T(P|k,k'')G(P/2+k'')G(P/2-k'')V(P|k'',k').
\end{equation}
Here $P\equiv(\Omega,{\bf P})$ is the total 4-momentum of two fermions, 
$k$ and $k'$ are relative 4-momenta, and ${\cal V}$ is the four-dimensional 
volume (with an imaginary time dimension).  
By reducing $G_2$ to a functional of $G$, one breaks the infinite hierarchy 
of equations for $n$-particle Green's functions because the resulting 
approximate Dyson equation $G=G^{(0)} + G\Sigma G^{(0)}$ 
contains no higher-order Green's functions
[Fig.~\ref{graphs}(c)]:
\begin{equation}
\label{Sigma}
\Sigma(k)
= -\frac{i}{\cal V}\sum_{P}
T(P|P/2-k,P/2-k)G(P-k),
\end{equation}
Equations \ref{T} and \ref{Sigma} form a 
closed set of equations that could be solved, at least in principle.
\cite{Micnas}

The use of full fermion propagators (instead of bare ones)
in these equations is the main difference from the earlier approach 
of Thouless,\cite{Thouless} equivalent to the BCS theory.  In addition 
to having conservation laws built in,\cite{Baym} the 
self-consistent approach is a step towards including the 
influence of Gaussian fluctuations on the phase transition.  Near the 
transition temperature, fermions start to feel the presence of the emerging
bosonic excitations, which in turn influences the process of pair formation.
Dressing of the fermionic propagators in Eq.~\ref{T} allows one to 
account for this feedback, at least to some extent.  The use of the dressed
fermion propagator in the self-energy equation (\ref{Sigma}) 
increases the number of terms in the perturbation series for $G(k)$,
which turns out to be important for treating properly the quantum nature of 
fluctuating Cooper pairs.  

\subsection{Propagator of a Cooper pair.}

Although the $T$ matrix depends on three independent momenta, 
the non-trivial dependence 
should be associated with the conserved total momentum $P$.  
It is convenient to write the interaction vertex formally as 
[Fig.~\ref{graphs}(d)]
\begin{equation}
\label{factorization}
V(P|k,k') = u^\dagger(k)D^{(0)}(P)u(k')
\equiv \int u^*(k|x)D^{(0)}(P|x,x')u(k'|x')\ d^4x\,d^4x',
\end{equation}
where $x$ and $x'$ are some relative coordinates.  
One well-known example of such factorization is the case of a separable 
instanteneous potential $V(P,k,k')=gv({\bf k})v({\bf k'})$.  This formalism
is particularly convenient for lattice models with finite-range instanteneous 
interactions.  In the case of the $t$-$J$ model, for instance,
\begin{equation}
V(P,k,k') = \left(
\begin{array}{ccc}
1 & \cos{k_x} & \cos{k_y}
\end{array}
\right)
\left(
\begin{array}{ccc}
+\infty & 0 & 0\\
0 & -J & 0\\
0 & 0 & -J
\end{array}
\right)
\left(
\begin{array}{c}
1\\ 
\cos{k_x'}\\ 
\cos{k_y'}
\end{array}
\right)
\end{equation}
in the singlet channel and zero in the triplet one.  
The infinite on-site repulsion imposes the constraint of no double 
occupancy.  Phonon-mediated attraction can also be written in the 
form (\ref{factorization}) with $u(k|x)=e^{i{\bf k}\cdot{\bf x}-i\omega t}$.

The $T$ matrix can now be found using the {\em ansatz} $T(P|k,k') = 
u^\dagger(k)D(P)u(k')$ [Fig.~\ref{graphs}(e)].  The new matrix $D(P|x,x')$ 
satisfies the Dyson equation $D = D^{(0)} + D\Pi D^{(0)}$ 
with the polarization matrix 
\begin{equation}
\label{Pi(P)}
\Pi(P) = \frac{i}{\cal V}\sum_{k}u(k)u^\dagger(k)G(P/2+k)G(P/2-k).
\end{equation}
Clearly, $D(P|x,x')$, which contains all the non-trivial information about 
the $T$ matrix, can be regarded as the bosonic propagator of a Cooper pair
whose internal structure is described by the dependence on the relative
coordinates $x$ and $x'$.  

The factorization procedure described above allows one to write the 
Bethe-Salpeter equation for the $T$ matrix (\ref{T})
in the form reminiscent of the Dyson equation for a 
doubly charged bosonic particle --- see Fig.~\ref{graphs}(f) --- without 
introducing spurious degrees of freedom.  The Thouless criterion
for pair condensation, $T(P)=\infty$ at $P=0$, simply means that 
the quasiparticles with ${\bf P}=0$ have zero excitation energy.  

\section{Fermion propagator in the $T$-matrix approximation.}
\label{fermions}

The fermion self-energy (\ref{Sigma}) can now be rewritten using the boson
propagator $D(P)$ [Fig.~\ref{graphs}(g)]:
\begin{equation}
\label{Sigma bosonized}
\Sigma(k) = -\frac{i}{\cal V}\sum_{P}u^\dagger(P/2-k)D(P)u(P/2-k)G(P-k).
\end{equation}
This diagram illustrates
two important points about the nature of the $T$-matrix approximation.  
(1) The irreversible decay of a fermionic quasiparticle is determined 
by the density of states ``hole + Cooper pair'' at a given 
4-momentum.  Note that both outgoing lines are dressed and thus 
represent actual excitations of the system.  
This should be contrasted to the functional-integration 
approach at the usual one-loop level\cite{Randeria,Pistolesi,Loktev}
where the fermion self-energy is expressed in terms of the {\em bare} 
fermion propagator.  We will return to this point later in 
Sec.~\ref{Gaussian fluctuations}.
(2) For a slowly fluctuating pairing field, the sum in 
Eq.~\ref{Sigma bosonized} is dominated by the region near $P=0$, so that 
a fermion with 4-momentum $k$ is coupled mostly to the hole with the same 
4-momentum.  By replacing $G(P-k)$ with $G(-k)$, we obtain 
\begin{equation}
\label{slow Sigma}
\Sigma(k) \approx - |\Delta(k)|^2 G(-k) 
\equiv \frac{|\Delta(k)|^2}{\Sigma(-k)-\left[G^{(0)}(k)\right]^{-1}},
\end{equation}
where 
\begin{equation}
|\Delta(k)|^2 \equiv
\frac{i}{\cal V}\sum_{P}u^\dagger(P/2-k)D(P)u(P/2-k)
\end{equation}
The approximation made here may appear rather crude.  Essentially, 
the scattering of a fermion into a {\em continuum} of hole states is replaced 
by the coupling to a {\em single} hole state with the 
same 4-momentum, a situation reminiscent of the BCS superconductor.  
If a bare propagator were used for the hole, the resulting 
Bogoliubov quasiparticles would be stable.  However, the use of a full 
propagator $G(-k)$ already makes the lifetime finite.   
In this case, neglecting temporal and spatial fluctuations of the pairing 
field does not look so bad, especially in a low-dimensional system.  

The approximate equation for the self-energy (\ref{slow Sigma}) can be
readily solved.\cite{Schmid}  Iterating it once 
results in a quadratic equation for $\Sigma(k)$.  Since
$|\Delta(k)|^2=|\Delta(-k)|^2$, the dressed propagator is
\begin{equation}
G(k) = \frac{2G^{(0)}(k)}{1+\sqrt{1+4|\Delta(k)|^2G^{(0)}(k)G^{(0)}(-k)}}.
\end{equation}
The branch of the square root is fixed by the requirement that 
$G(k) \to G^{(0)}(k)$ as $|\Delta(k)|\to 0$.  

In the simplest case of a short-range instanteneous attractive potential,
$\Delta(k)$ can be replaced by a constant and we have 
\begin{equation}
\label{dressed instantaneous}
G(\omega,{\bf k}) = \frac{1}{\omega - \epsilon_{\bf k}}
\left(\frac{1}{2}
+\sqrt{\frac{1}{4}-\frac{|\Delta|^2}{\omega^2-\epsilon_{\bf k}^2}}\right)^{-1},
\end{equation}
where $\epsilon_{\bf k}$ is the bare fermion energy.
It is readily seen that the spectral weight of the dressed fermion 
is distributed over two finite regions $\epsilon_{\bf k}^2<\omega^2<
\epsilon_{\bf k}^2+4|\Delta|^2$ as follows:
\begin{equation}
\label{weight in T}
{\cal A}(\omega,{\bf k}) = \frac{|\omega+\epsilon_{\bf k}|}{2\pi|\Delta|^2}
\sqrt{\frac{\epsilon_{\bf k}^2+4|\Delta|^2-\omega^2}
{\omega^2-\epsilon_{\bf k}^2}}.
\end{equation}
This distribution is reminiscent of the smeared BCS peaks at 
$\omega^2=\epsilon_{\bf k}^2+|\Delta|^2$.  
In fact, the ratio of the spectral weights 
(\ref{weight in T}) at $\omega=\pm\sqrt{\epsilon_{\bf k}^2+|\Delta|^2}$ 
is the same as in the BCS case.  

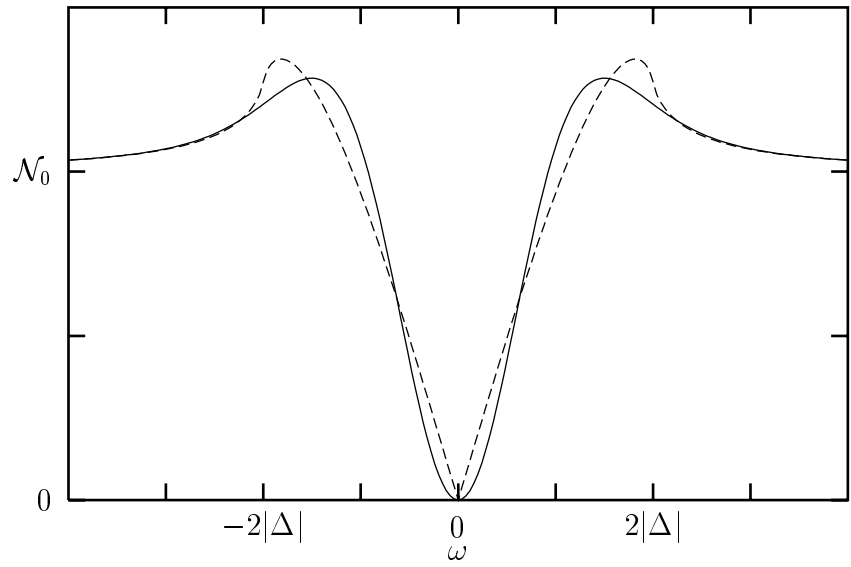
\begin{figure}
\caption{Pseudogap in the fermion density of states in the limit of slow 
pairing fluctuations.  Dashed line: $T$-matrix approximation.  Solid line: 
exact model with Gaussian fluctuations.}
\label{pseudogap}
\end{figure}

Since the spectral weight is pushed away from the Fermi surface, 
a pseudogap opens up in the total density of states ${\cal N}(\omega)$
(Fig.~\ref{pseudogap}):
\begin{equation}
{\cal N}(\omega) = {\cal N}_0\;f(\omega/2|\Delta|),
\end{equation}
where ${\cal N}_0$ is the density of states in the free case and the 
function $f(x)$ can be expressed in terms of complete elliptic integrals:
\begin{equation}
f(x) = 
\left\{
\begin{array}{l}
(4/\pi)x{\bf E}(x) \mbox{ for } 0<x<1,\\
(4/\pi)[{\bf K}(1/x)-{\bf D}(1/x)] \mbox{ for } x>1.
\end{array}
\right.
\end{equation}
It vanishes linearly for small values of $x$ and approaches 1 as 
$x\to \infty$.    

\section{Ideal fermions interacting with an independent pairing field.}
\label{Gaussian fluctuations}

The $T$-matrix approximation amounts to summing up an infinite
number of terms of the perturbation series.  Yet an even ``greater'' 
infinity of terms is left out.  In such a case, when approximations 
are not easily controlled, it is desirable to 
check the result using some exactly solvable model.  

Consider a free fermion gas interacting with a classical external 
pairing field, which is represented by the action term
\begin{equation}
\label{S_int}
S_{\rm int} = \int [\Delta(x,x') \psi^\dagger(x)\psi^\dagger(x')
+ \mbox{H.c.}]d^4x\; d^4x'.
\end{equation}
The field $\Delta(x,x')$ can be interpreted as the wavefunction of a 
Cooper pair with its constituent fermions at $x$ and $x'$; the interaction 
term above removes two free fermions and forms a bound state.  

The $U(1)$ symmetry, related to the conservation of charge,  
makes the phase of the complex
field $\Delta$ unobservable.  When pair formation lowers the 
energy of the system, the density of pairs $|\Delta|^2$ may become 
large.  There are two distinct possibilities in this case.  The first one
occurs when the $U(1)$ symmetry is sponateously broken and $\Delta$ 
chooses some direction in the complex plane with the well-known results
(the BCS superconductor with a gap and a Bogoliubov sound mode).
The second possibility 
is a symmetric phase with a fluctuating field $\Delta$.  The 
quantum nature of these fluctuations leads to a non-trivial
effect: a significant broadening of the fermion spectral weight.  

Quantum fluctuations in this model can be implemented 
by using an ensemble of pairing fields $\Delta$ with a symmetric
distribution.  Although every {\em single} measurement of $\Delta$ produces
a definite non-zero result, $\langle\Delta\rangle=0$ upon averaging over 
a long series of such measurements.  This procedure is 
Feynman's way of quantizing a classical system.\cite{Feynman}  
A closely related exactly solvable model for the 
Peierls gap in quasi-one-dimensional metals has been 
discussed by McKenzie\cite{McKenzie} following the analytic solution of 
Sadovskii.\cite{Sadovskii}

\subsection{Broken phase.}
\label{broken phase}

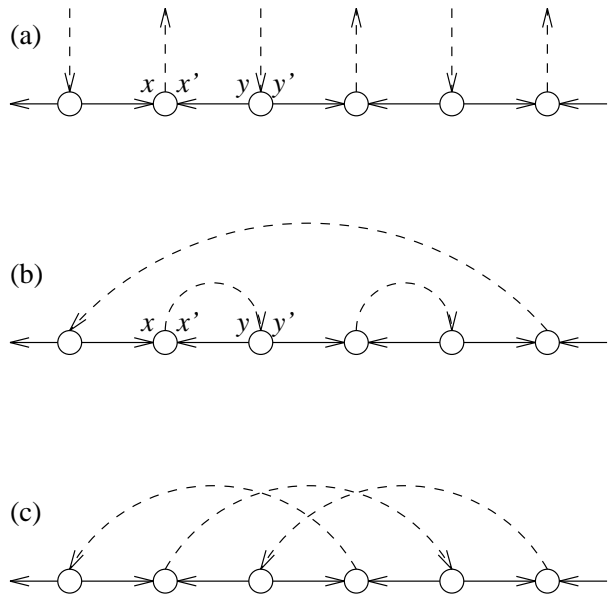
\begin{figure}
\caption{Feynman diagrams for the exact model with a classical pairing field.
Solid lines are bare fermion propagators.  (a) A diagram of order $n=3$. 
Dashed lines 
with circles are pairing fields $\Delta(x,x')$ and $\Delta^*(y,y')$.  
(b)-(c) Examples of diagrams generated from (a) 
after averaging over a Gaussian ensemble of pairing fields.  Dashed lines
are now pairwise averages $\langle\Delta(x,x')\Delta^*(y,y')\rangle$.}
\label{gauss}
\end{figure}

A generic diagram of the perturbation series for the fermion propagator 
contains $n$ incoming and $n$ outgoing dashed lines representing the 
pairing field [Fig.~\ref{gauss}(a)].  
If $\langle\Delta(x,x')\rangle\neq 0$, there is also the anomalous fermion
propagator, which gives rise to the Meissner effect and superconductivity.  
Consider the extreme case when there are 
no fluctuations in the ensemble: $\Delta(x,x')=\langle\Delta(x,x')\rangle$.
The dressed fermion propagator is simply a geometrical series with a single 
term in each order.  For an instanteneous short-range pairing field 
$\langle\Delta(x,x')\rangle=\overline{\Delta}\delta(x-x')$ (uniform for 
simplicity), 
\begin{equation}
\label{broken}
G(k)=\frac{1}{\omega-\epsilon_{\bf k}}\sum_{n=0}^{\infty}
\left(\frac{\left|\overline{\Delta}\right|^2}
{\omega^2-\epsilon_{\bf k}^2}\right)^n.
\end{equation}
Although the geometrical series 
$\sum_{n=0}^{\infty}z^n$
diverges when $|z|>1$, its analytic continuation, the function 
$(1-z)^{-1}$, has only one singular point $z=1$.  The BCS result is recovered:
\begin{equation}
\label{BCS G}
G(k)=\frac{\omega+\epsilon_{\bf k}}
{\omega^2-\epsilon_{\bf k}^2-\left|\overline{\Delta}\right|^2}
\end{equation}
The divergence of the perturbation series (\ref{broken}) in the range 
$\epsilon_{\bf k}^2<\omega^2<\epsilon_{\bf k}^2+
\left|\overline{\Delta}\right|^2$ 
is associated with the transfer of the 
pole of the propagator from $\omega=\epsilon_{\bf k}$ to the new locations at 
$\omega^2=\epsilon_{\bf k}^2+\left|\overline{\Delta}\right|^2$.  

\subsection{Symmetric phase with quantum fluctuations.}
\label{frozen fluctuations}

The structure of the perturbation theory changes considerably 
in the symmetric phase.  Assuming that the statistical distribution 
of the pairing fields is Gaussian, we can write any average that involves 
a product of $n$ fields $\Delta(x_m,x_m')\equiv\Delta_m$ and 
$n$ fields $\Delta^*(y_m,y_m')\equiv\Delta^*_m$
as a sum over all possible products of pairwise averages:
\begin{equation}
\langle\Delta^*_1\Delta_1\Delta^*_2\Delta_2\ldots\Delta^*_n\Delta_n\rangle
=\sum_{P}
\langle\Delta^*_1\Delta_{P1}\rangle
\langle\Delta^*_2\Delta_{P2}\rangle
\ldots
\langle\Delta^*_n\Delta_{Pn}\rangle,
\end{equation}
where $P$ represents a permutation of the numbers $1\ldots n$.
Upon introducing a dashed line for each pair average
$\langle\Delta^*_i\Delta_j\rangle$, we find 
$n!$ different diagrams with $n$ such
lines [Fig.~\ref{gauss}(b) and (c)].  Again, take the example of a uniform
instantaneous short-range pairing field:
\begin{equation}
\langle\Delta^*(y,y')\Delta(x,x')\rangle = \overline{|\Delta|^2}
\delta(x-x')\delta(y-y').
\end{equation}
Each {\em single} field breaks the phase symmetry and produces 
the effect described above.  Averaging over the ensemble, 
however, restores this symmetry; the perturbation series now contains 
$n!$ identical terms in the $n$-th order:
\begin{equation}
\label{restored}
G(k)=\frac{1}{\omega-\epsilon_{\bf k}}\sum_{n=0}^{\infty}
n!\left(\frac{\overline{|\Delta|^2}}{\omega^2-\epsilon_{\bf k}^2}\right)^n.
\end{equation}
Clearly, the result is different from (\ref{broken}).
Expression (\ref{restored}) was obtained in Ref.~\ref{Ren} 
in connection with the pseudogap in the boson-fermion model --
as a leading term in two dimensions.  

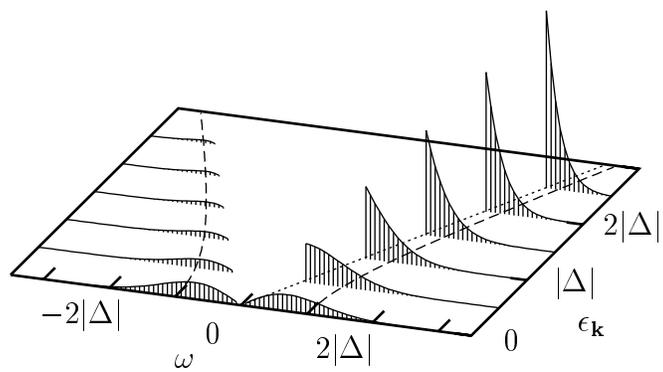
\begin{figure}
\caption{Distribution of the fermion spectral weight 
${\cal A}(\omega,{\bf k})$ in the model with Gaussian pairing fluctuations.  
$\epsilon_{\bf k}$ is the bare energy of a fermion, $|\Delta|$ is the 
r.m.s. fluctuation of the pairing field.  Also shown are positions of the
fermion poles in a non-interacting fermion system (dotted line) and a BCS
superconductor with the gap $|\Delta|$ (dashed line).}
\label{weight}
\end{figure}

Although the series 
\begin{equation}
\sum_{n=0}^\infty n!z^n
\end{equation}
diverges for any $z\neq 0$, it represents 
an asymptotic Taylor expansion of the function
\begin{equation}
f(z) = \int_0^\infty\frac{e^{-t}dt}{1-zt}. 
\end{equation}
The divergence of the series is related to the nonanalyticity of $f(z)$ on the 
positive side of the real axis, where it has a cut.  
Rather then working with a series that converges only in the trivial
case $z=0$, we can sum up the diagrams for a single uniform field $\Delta$ --
as in Sec.~\ref{broken phase} -- and then average the result over a Gaussian 
ensemble with the weight 
$\exp{\left(-|\Delta|^2/\overline{|\Delta|^2}\right)}d\Delta^*d\Delta$:
\begin{equation}
G(k)  
= \frac{1}{\omega-\epsilon_{\bf k}}
\int\frac{e^{-t}dt}{1-\frac{\overline{|\Delta|^2}}
{\omega^2-\epsilon_{\bf k}^2}t},
\end{equation}
which indeed has (\ref{restored}) as an asymptotic expansion.  
The quasiparticle spectral weight, concentrated at $\omega=\epsilon_{\bf k}$ 
in the free case, is now spilled into the region 
$\omega^2>\epsilon_{\bf k}^2$ (Fig.~\ref{weight}):
\begin{equation}
{\cal A}(\omega,{\bf k})
= \frac{|\omega+\epsilon_{\bf k}|}{\overline{|\Delta|^2}}
\exp{\left(-\frac{\omega^2-\epsilon_{\bf k}^2}{\overline{|\Delta|^2}}\right)}, 
\ \ \omega^2>\epsilon_{\bf k}^2.
\end{equation}

The density of states exhibits a pseudogap (Fig.~\ref{pseudogap}), 
somewhat more pronounced than it was found in the $T$-matrix approximation: 
it now vanishes as $\omega^2$ near the Fermi level.  

The results of this section can be understood in a simple intuitive way.  
Even though we have suppressed spatial and temporal fluctuations of the 
pairing field, its strength does not yet have a definite value; the fermion 
pole, instead of being shifted to a definite new position, is 
smeared into a cut along some part of the real axis.  Nevertheless, 
the existence of a scale $\overline{|\Delta|^2}$ in the field distribution 
results in the fermion spectral function reminiscent of the BCS one.  

\subsection{Relation to the $T$-matrix approximation.}

The $T$-matrix equation for the fermion self-energy (\ref{slow Sigma}),
which includes a {\em dressed} fermion propagator, is equivalent to 
partial summation of the diagrams in Sec.~\ref{frozen fluctuations}: 
only those without intersecting dashed lines are included.  For instance, 
diagram (b) in Fig.~\ref{gauss} is present, while (c) is left out.  
The number of diagrams in the $n$-th order satisfies the recursion relation
\begin{equation}
\label{recursion}
C_{n+1} = \sum_{m=0}^{n}C_m C_{n-m} 
\end{equation}
($C_0=1$ by definition) with the solution
\begin{equation}
C_n = \prod_{m=1}^{n}\left(4-\frac{6}{m+1}\right),
\end{equation}
which grows as $4^n$ for a large $n$.  Therefore, the series 
\begin{equation}
\label{my Taylor}
\sum_{n=0}^\infty C_n z^n 
\end{equation}
has the radius of convergence equal to 1/4 and the Green's function
\begin{equation}
\label{disentangled series}
G(k) = \frac{1}{\omega-\epsilon_{\bf k}}\sum_{n=0}^\infty C_n 
\left(\frac{\overline{|\Delta|^2}}{\omega^2-\epsilon_{\bf k}^2}\right)^n
\end{equation}
is analytical (hence real) when 
$\omega^2>\epsilon_{\bf k}^2+4\overline{|\Delta|^2}$.  
The spectral weight is contained in the finite region
$\epsilon_{\bf k}^2<\omega^2<\epsilon_{\bf k}^2+4\overline{|\Delta|^2}$.

The above discussion shows that the $T$-matrix approximation 
includes, at least partially, effects
of quantum fluctuations of the pairing field on the fermion propagator.
This is achieved through the use of a
dressed fermion propagator in the equation for the fermion 
self-energy.  The number of terms in the resulting perturbation series grows 
quickly enough to make the result non-analytical in a larger region 
than in the BCS broken phase.  Inclusion of a greater number of terms 
(the full series) only slightly enhances this effect.  
This should be contrasted to the situation when a {\em bare} fermion 
propagator is used in the fermion self-energy (\ref{slow Sigma}).  Since 
there is only one graph in each order of the perturbation series
for the propagator $G(k)$, such a treatment essentially amounts to 
throwing away the quantum nature of Cooper pairs.  No broadening is 
then found in the limit of slow fluctuations.  

\section{Cooper pairs in a pseudogap.}
\label{bosons}

\subsection{The propagator of a Cooper pair.}
\label{boson propagator}

With much of the spectral weight removed from the vicinity of the Fermi 
surface, low-energy bosonic excitations may represent a propagating 
mode.  In this respect, such a normal state resembles a BCS 
superconductor just below $T_c$, where the opening of a true gap inhibits 
the decay of pairs.\cite{dissipation}  Therefore, a reasonable estimate of 
the boson energy spectrum can be obtained from Eq.~\ref{Pi(P)} by using 
the dressed fermion propagator in the BCS form (\ref{BCS G}), which 
constitutes the two-pole {\em ansatz} of Ref.~\ref{Micnas}.  Inclusion of 
fermion lifetime effects makes Cooper pairs unstable but is not expected
to produce qualitative changes in the energy spectrum.
For a short-range instanteneous attraction, $D^{(0)}(P)=g<0$, 
\begin{equation}
\label{inverse D}
D^{-1}(P) = g^{-1}+\frac{i}{\cal V}\sum_{k}G(k)G(P-k),
\end{equation}
with 
\begin{equation}
\label{approximate G}
G(k)\equiv G(\omega,{\bf k})=\frac{\omega+\epsilon_{\bf k}}
{\omega^2-\epsilon_{\bf k}^2-\overline{|\Delta|^2}}.
\end{equation}

At this level of approximation (no boson decay because of a full fermion gap), 
$D^{-1}(\Omega,{\bf P})$, analytically continued from Matsubara frequencies
to the rest of the $\Omega$ plane, can be expanded in powers of $\Omega$ 
and $\bf P$:
\begin{equation}
\label{D-inverse}
D^{-1}(\Omega,{\bf P}) = D^{-1}(0,0) 
+ \frac{\partial D^{-1}(0,0)}{\partial \Omega^2}\Omega^2 
+ \frac{\partial D^{-1}(0,0)}{\partial {\bf P}^2}{\bf P}^2 
+ \ldots
\end{equation}
or 
\begin{equation}
\label{pseudorelativistic}
D(\Omega,{\bf P}) \approx \frac{Z}{\Omega^2-{\bf P}^2s^2 - M^2 s^4}.
\end{equation}
The mass term appears because the Thouless criterion no longer applies:
$D^{-1}(0,0)\neq 0$ in the normal state.  The existence of {\em two} 
Cooper-pair poles at a given momentum $\bf P$, at the frequencies 
$\Omega=\pm\sqrt{M^2 s^4 + {\bf P}^2 s^2} \equiv \pm E_{\bf P}$, 
reflects the fact that correlated propagation is possible not only 
for two fermionic particles, but also for two holes.  If a pair 
is made out of fermionic excitations just around the Fermi surface, 
where the density of states is the same for particles and holes, 
the residues of the two poles are equal (up to a sign).  A varying
density of states will make the two poles less symmetric, both in terms 
of their residues and positions, but still low-energy Cooper pairs will have 
two poles.  

This trend can be clearly seen in the numerically obtained 
two-particle density of states plotted in Fig.~4 of Ref.~\ref{Micnas}.  
Low-momentum pairs have two somewhat asymmetric peaks dispersing with 
momentum.  The asymmetry is related to the instanteneous nature 
of the interaction in the attractive Hubbard model: the number of
particle states involved in the formation of a pair greatly exceeds 
that of hole states at low fermion densities.  
No holes contribute to the formation of pairs with higher momenta,
so that there is no pole at negative frequencies in this case.  Such pairs, 
however, represent fast fluctuations of the pairing field and are irrelevant to
the formation of the pseudogap.  The numerical data also show
an increasing lifetime of Cooper pairs as ${\bf P}\to 0$, reflecting a 
stronger depletion of the fermion density of states near the Fermi energy.

When fermionic excitations across the gap are frozen out, the Cooper-pair
propagator at zero momentum is
\begin{equation}
\label{inverse D(Omega,0)}
D^{-1}(\Omega,0) = g^{-1} -\int_{-\infty}^{\infty}{\cal N}(\epsilon)d\epsilon
\frac{\epsilon^2+\tilde{\epsilon}^2+\epsilon\Omega}
{\tilde{\epsilon}(\Omega^2-4\tilde{\epsilon}^2)}
\end{equation}
where $\tilde{\epsilon}\equiv\sqrt{\epsilon^2+\overline{|\Delta|^2}}$.
With the exception of the zeroth-order term $D^{-1}(0,0)\equiv -Z^{-1}M^2s^4$, 
coefficients of the Taylor series (in powers of $\Omega$) are insensitive 
to large-$\epsilon$ behavior of the density of states, and are thus
determined by the lower energy scale, {\em i.e.,} the gap width $|\Delta|$,
provided that variations of the bare density of states are small on that 
energy scale.  
In this weak-coupling limit, the terms odd in $\Omega$ vanish because 
${\cal N}(\epsilon)\epsilon$ is an odd function near $\epsilon=0$.   
Quartic and higher-order terms can be dropped as long as 
$\Omega^2\ll\overline{|\Delta|^2}$.  Then
\begin{equation}
D^{-1}(\Omega,0) \approx 
Z^{-1}(\Omega^2 - M^2s^4),
\end{equation}
where $Z^{-1}$ is of order ${\cal N}_0/\overline{|\Delta|^2}$.  More exactly,
\begin{equation}
\label{Z}
Z = 12\overline{|\Delta|^2}/{\cal N}_0.
\end{equation}

Expansion in powers of $\bf P$ allows one to determine the characteristic
speed of bosons $s$, which is expected to be smaller than the Fermi velocity
$v$ on physical grounds.  In the weak-coupling limit, we recover the result
of Bogoliubov, $s^2=v^2/d$ in $d$ dimensions.
This is roughly consistent with the dispersion of Cooper-pair poles 
at low momenta inferred from the data of Ref.~\ref{Micnas}.  

At a given temperature, two as yet undetermined low-energy 
parameters of the boson field -- the mass $M$ and the pole 
residue $Z$ -- control the intensity of fermion scattering by the 
pairing field and thus determine the (also unknown) fermion energy gap,
as discussed in some detail below.  In addition, Eq.~\ref{Z} relates
$Z$ to the energy gap.  Using this information, it should be possible to 
determine two out of the three unknown parameters.  In the 
following section, the mass $M$ is found as a function of temperature.  

\subsection{Evaluation of the fermion self-energy}
\label{self-energy}

We now return to the analysis of the fermion self-energy 
(\ref{Sigma bosonized}) using the approximate boson propagator 
(\ref{pseudorelativistic}).  The following expression for the fermion 
self-energy results:
\begin{equation}
\Sigma(\omega,{\bf k}) \approx \int \frac{d^d{\bf P}}{(2\pi)^d}
\frac{1}{2\pi i}
\oint \frac{d\Omega}{e^{\beta\Omega}-1}
\frac{Z}{\Omega^2-E_{\bf P}^2}G(\Omega-\omega,{\bf P-k}).
\end{equation}
The sum over the bosonic Matsubara frequencies $\Omega$ has been 
converted into an integral around the poles of $(e^{\beta\Omega}-1)^{-1}$;
$E_{\bf P} \equiv \sqrt{M^2 s^4 + {\bf P}^2 s^2}>0$ is the boson energy.  
It is convenient to write the fermion propagator in terms of its spectral 
weight:
\begin{equation}
G(\omega,{\bf k}) = \int_{-\infty}^{+\infty} 
\frac{{\cal A}(\epsilon',{\bf k})d\epsilon'}
{\omega-\epsilon'}.
\end{equation}
After deforming the integration contour in the standard way, one finds
the self-energy as a sum of two terms, which correspond to emission ($+$) 
and absorption ($-$) of a Cooper pair:
\begin{equation}
\label{Sigma via A}
\Sigma^{(\pm)}(\omega, {\bf k}) = Z\int \frac{d^d{\bf P}}{(2\pi)^d}
\frac{1}{2E_{\bf P}}\int {\cal A}(\epsilon',{\bf P-k})\;d\epsilon'
\frac{N(E_{\bf P})+n(\pm\epsilon')}{\omega+\epsilon'\mp E_{\bf P}}.
\end{equation}
The part proportional to the boson occupation number $N(E_{\bf P})$ 
can be regarded as induced emission or absorption with the rate 
proportional to the spectral intensity of the pairing field 
$N(E_{\bf P})/E_{\bf P}$.  The fermionic term $n(\pm\epsilon')$ is responsible
for ``spontaneous'' processes.  

If the pseudogap formation is governed by low-energy modes, the induced
term is dominant since $N(E_{\bf P})\gg n(\pm\epsilon')$ at low energies.
The rate of spontaneous emission and absorption of pairs is estimated 
in the Appendix; it can be neglected in $d=2$ dimensions.  
Upon shifting the integration variable 
$\epsilon'\mp E_{\bf P}\to\epsilon'$, 
\begin{equation}
\label{Sigma bosonic}
\Sigma_{\rm ind}^{(\pm)}(\omega, {\bf k}) = Z\int \frac{d^d{\bf P}}{(2\pi)^d}
\frac{N(E_{\bf P})}{2E_{\bf P}}
\int \frac{{\cal A}(\epsilon'\pm E_{\bf P},{\bf P-k})\;d\epsilon'}
{\omega+\epsilon'}
\end{equation}
Bosons are restricted to low-energy modes with $E_{\bf P}$ of order or less 
than 
$k_B T$.  In the case when the fermion spectral weight is distributed over 
a wider interval of energies, the variation of the fermion spectral 
weight on the energy scale of $k_B T$ can be neglected.  For the same reason,
we can neglect the variation of the fermion momentum in the integrand,
provided that fermion and boson velocities are of the same order
(recall that $s^2=v^2/d$).  The integration over $\epsilon'$ yields 
the fermion propagator $G(-\omega,{\bf -k})$ and the self-energy has the 
form inferred previously (\ref{slow Sigma}):
\begin{equation}
\label{Sigma bosonic smoothed}
\Sigma_{\rm ind}^{(+)}(\omega, {\bf k}) 
+ \Sigma_{\rm ind}^{(-)}(\omega, {\bf k}) 
\approx -\overline{|\Delta|^2} G(-\omega,-{\bf k}),
\end{equation}
where the average fluctuation (intensity) of the pairing field is
\begin{equation}
\label{fluctuation of Delta}
\overline{|\Delta|^2} = Z\int \frac{d^d{\bf P}}{(2\pi)^d}
\frac{N(E_{\bf P})}{E_{\bf P}}.
\end{equation}

Essentially the same result 
can be obtained by considering a classical field $\Delta(t,{\bf r})$ 
with the Lagrangian density determined by (\ref{pseudorelativistic}),
\begin{equation}
\label{Lagrangian}
{\cal L}(t,{\bf r}) = 
Z^{-1}\left[|\partial \Delta(t,{\bf r})/\partial t|^2 
- |\nabla\Delta(t,{\bf r})|^2 - Ms^2|\Delta(t,{\bf r})|^2\right],
\end{equation}
in thermal equilibrium.  Application of the equipartition theorem yields
\begin{equation}
\label{classical fluctuation of Delta}
\overline{|\Delta|^2} = Z\int \frac{d^d{\bf P}}{(2\pi)^d}
\frac{k_B T}{E_{\bf P}^2},
\end{equation}
which is the classical analogue of Eq.~\ref{fluctuation of Delta}.  

\subsection{Condensation temperature.}
\label{Condensation temperature}

By substituting the value of the boson residue (\ref{Z})
into Eq.~\ref{fluctuation of Delta}, 
we obtain the self-consistency condition mentioned above.  Remarkably, 
the width of the pseudogap cancels out and the resulting equation implicitly
determines the boson mass as a function of temperature:
\begin{equation}
\int\frac{d^d{\bf P}}{(2\pi)^d}\frac{N(E_{\bf P})}{E_{\bf P}} 
= \frac{{\cal N}_0}{12},
\end{equation}
where $E_{\bf P} = \sqrt{M^2 s^4 + {\bf P}^2 s^2}$, 
$N(E)=(e^{\beta E}-1)^{-1}$ is the boson occupation number, 
and ${\cal N}_0$ is the density of bare fermion 
states.  It is immediately obvious that there is no Bose condensation in 
$d=2$ dimensions: for $M=0$, the integral diverges at the lower
limit.  In $d=3$ dimensions, this equation predicts a condensation 
temperature $k_B T_c$ of order $\epsilon_F$.  This result should not be
taken at face value because it was derived for a weakly attractive 
degenerate fermion gas.  (Also, as discussed in the Appendix, the absence of 
a low energy scale implies that pairing fluctuations are not slow.)
Nevertheless, the existence of such a high
temperature scale is justified if one considers the limits of weak 
and strong attraction between fermions.  In the weak-coupling limit, 
there are no pairs at the Fermi temperature, 
they form at a much lower temperature $k_B T_0\ll\epsilon_F$; therefore,
the long-overdue condensation occurs immediately, which explains 
why the mean-field BCS approach works so well.  In the 
opposite limit of local pairs (turned into hard-core bosons), 
condensation indeed occurs when the gas becomes degenerate.  

Next, we estimate the condensation temperature
in $d=2+\varepsilon$ dimensions, which may be relevant to highly anisotropic 
cuprate superconductors.  Recalling the asymptotic behavior of the Riemann 
zeta-function $\zeta(1+\varepsilon)\sim 1/\varepsilon$ we obtain
\begin{equation}
\label{T_c near d=2}
k_B T_c \sim \varepsilon\epsilon_F/6
\end{equation}
when $\varepsilon\ll 1$.
For comparison, the condensation temperature of a Bose gas made of very small 
fermionic pairs is only higher by a factor $3/2$.  (In both cases, 
$\epsilon_F$ is the Fermi energy of the ideal fermion gas with the same 
mass and density of particles.\cite{MRR})  It is thus clear that the 
condensation 
of Cooper pairs in the presence of a pseudogap occurs well below the 
fermion degeneracy temperature and a clear separation of energy scales 
(\ref{separate scales}) is possible for moderately weak 
attraction strengths.  If the attraction is strong,
we end up with tightly bound pairs, in the local boson limit.  In the case 
of a very weak attraction, the BCS pair formation temperature $T_0$ is lower 
than $T_c$ of Eq.~\ref{T_c near d=2}; the BCS model takes over.  

\section{Conclusion}
\label{tired}

I have considered the interaction between fermions and Gaussian pairing 
fluctuations (Cooper pairs without self-interaction).  A well-defined 
pseudogap regime is found in $d=2+\varepsilon$ spatial dimensions,
when the condensation temperature is much lower than the BCS mean-field one.
The quantum character of Cooper pairs, related to the 
unobservable nature of the pair wavefunction $\Delta(x,x')$, leads to 
considerable broadening of the Bogoliubov quasiparticle peaks near
the Fermi surface.  Even when non-uniform configurations 
of the field $\Delta(x,x')$ are suppressed at low temperatures, 
fluctuations of its {\em amplitude} blur the strength of the pairing field,
making the position of the quasiparticle poles at 
$\omega^2=\epsilon_{\bf k}^2+|\Delta(k)|^2$ uncertain.  

A well-pronounced suppression of the fermion density of states 
near zero energy makes room for long-lived pair states inside the gap.
At low momenta and frequencies, they represent a Bogoliubov sound-like mode 
with a non-zero mass.  The mass is determined from a self-consistency 
condition that links the width of the pseudogap to the mean fluctuation of 
the pairing amplitude.  Zero mass signals the onset of Bose condensation.  
In $d=2+\varepsilon$ dimensions, the condensation temperature scales 
as $\varepsilon\epsilon_F/6$, which is $2/3$ times the condensation 
temperature for the corresponding ideal Bose gas.  This may explain the doping 
dependence of $T_c$ in underdoped cuprates.  

Approximations made in this work are twofold.  First, 
an infinite hierarchy of equations for $n$-particle Green's 
functions is replaced with the self-consistent $T$-matrix approach
of Brueckner, which partially accounts for Gaussian pairing fluctuations.
By solving an exact model with an independent pairing field, I show 
that the self-consistent $T$-matrix approximation takes into account the
amplitude fluctuations of Cooper pairs in the normal state, a feature missing
in other approximate models.\cite{Randeria,Loktev,Emery}  Further 
approximations made in this paper are necessary to obtain an analytic
solution.  Firstly, thermally excited non-uniform 
configurations of the pairing field have little effect on the fermion
propagator, provided that
a clear separation of energy scales, $k_B T_c\ll |\Delta|$, exists.  
This appears to be the case in $d=2+\varepsilon$ dimensions, 
{\em i.e.,} for highly anisotropic, almost two-dimensional systems.  
Secondly, a residual density of fermionic states in the pseudogap, leading to 
the decay of Cooper pairs, is assumed to have negligible influence 
on the pair energy spectrum.  A numerical work using the 
self-consistent $T$-matrix approach\cite{Micnas} seems to indicate that 
this is a reasonable assumption: low-energy Cooper pairs represent a 
propagating mode with two Cooper-pair poles dispersing with the pair 
momentum.  In order to ascertain that propagating Cooper pairs in the 
pseudogap are not an artifact of the self-consistent $T$-matrix 
approximation, it is necessary to look for them in quantum Monte-Carlo 
simulations on two-dimensional lattices.  

There are many questions that remain open.  For instance, in the case of 
a superconducting order parameter with nodes, the density of 
fermion states is less strongly suppressed.  Will Cooper pairs
be stable enough to represent a propagating mode?  Also, what causes 
the strong broadening of fermion quasiparticle peaks above the 
temperature at which pairs are formed?  

\section*{Acknowledgments.}

The idea of this work has originated in conversations with Y.~J.~Uemura.
I thank A.~S.~Blaer, D.~M.~Eagles, T.~D.~Lee, R.~Micnas,
M.~Randeria, M.~V.~Sadovskii, and especially H.~C.~Ren for helpful 
discussions.  
This work has been supported in part by the U.~S.~Department of Energy 
(grant DE-FG02-92 ER40699) and by NEDO (Japan).  

\appendix
\section*{}
\label{slow appendix}

Spontaneous emission or absorption of Cooper pairs is not restricted 
to low boson momenta and thus represents the effect of fast fluctuations.
The emission or absorption rate depends on the phase space available to
the products of a fermion decay (a pair and a hole), which tends to 
be smaller in a lower number of dimensions $d$.  This indeed is confirmed
by a simple calculation below.  In $d=3$ dimensions, the imaginary part 
of $\Sigma_{\rm sp}(\omega,{\bf k})$ is linear in $\omega$ 
and thus leads to a renormalization of the quasiparticle spectral weight.
In $d=2$ dimensions, a constant imaginary part is introduced, which 
merely broadens the quasiparticle peak.  Besides, the existence
of a small parameter in the pseudogap regime (the ratio $|\Delta|/\epsilon_F$)
makes this additional broadening insignificant 
in $d=2+\epsilon$ dimensions.  

Let us estimate the impact of ``vacuum'' fluctuations on the 
self-energy of a fermion with momentum $\bf k$ and bare energy 
$\epsilon_{\bf k}=0$.  
Inasmuch a hole created in the decay can be far away from the 
Fermi surface, we can neglect the influence of pair scattering on its spectrum
and treate the hole as a stable particle with momentum $\bf -k' = k-P$ and 
energy 
\begin{equation}
-\epsilon_{\bf k'} 
= -\epsilon_{\bf k} + \frac{{\bf k\cdot P}}{m} -\frac{|{\bf P}|^2}{2m}
\approx |{\bf P}|v\cos{\theta} \equiv -\epsilon'(|{\bf P}|,\theta),
\end{equation}
neglecting the curvature of the energy surface $\epsilon_{\bf k}$; 
$\theta$ is the angle between $\bf k$ and $\bf P$.  The boson energy
is taken to be $E_{\bf P} = s |{\bf P}|\equiv E(|{\bf P}|)$.  Then
\begin{equation}
{\rm Im}\;\Sigma_{\rm sp}^{(\pm)}(\omega,{\bf k}) =
-Z\pi\int\frac{d^d{\bf P}}{(2\pi)^d}
\frac{n(\pm \epsilon')\delta(\omega+\epsilon'\mp E)}{2E}
\end{equation}
Energy conservation requires that 
\begin{equation}
E=\frac{\omega s}{s + v\cos{\theta}}.
\end{equation}
For definiteness, let us assume that the frequency of the incoming fermion 
$\omega>0$.  Then a pair can be emitted ($E>0$) when 
$\cos{\theta}>-s/v$ and absorbed ($E<0$) when $\cos{\theta}<-s/v$.  
Pair emission is accompanied by absorption of a second fermion 
with the energy in the range $\epsilon'>-v\omega/(v+s)\equiv\omega_+$.
The second fermion can, thus, be taken from {\em below} the Fermi 
surface and the process will take place even at low temperatures!
On the contrary, absorption of a pair requires emission of a second 
fermion with a negative energy $\epsilon'<-v\omega/(v-s)\equiv\omega_-$.  
At low temperatures, such states are occupied and pair absorption is 
suppressed.  

In $d=3$ dimensions, the element of phase space reduces to 
$E\;dE\;d\epsilon'/(2\pi)^2 vs^2$.  The integration over the boson energy 
$E$ is elementary because a $\delta$-function is present:
\begin{equation}
{\rm Im}\;\Sigma_{\rm sp}(\omega,{\bf k}) = -\frac{Z}{8\pi vs^2}
\left[
\int_{-\omega_+}^{\infty} n(\epsilon')d\epsilon'
+\int_{-\infty}^{-\omega_-}
n(-\epsilon')d\epsilon'
\right],
\end{equation}
The first term in brackets (emission of a pair) grows linearly 
with $\omega$.  By inserting $Z$ from (\ref{Z}), we obtain
\begin{equation}
{\rm Im}\;\Sigma_{\rm sp}(\omega,{\bf k}) \sim
-\frac{\pi\overline{|\Delta|^2}\omega}{\epsilon_F^2(1+1/\sqrt{3})}
\end{equation}
as $\omega\to+\infty$.  Thus, spontaneous pair emission leads to 
a significant renormalization of the fermion propagator 
(the ratio $\overline{|\Delta|^2}/\epsilon_F^2$ is not small for the pseudogap
regime in $d=3$) and cannot be neglected.  

In $d=2$ dimensions, pair emission has a smaller available phase space:
\begin{equation}
{\rm Im}\;\Sigma_{\rm sp}^{(+)}(\omega,{\bf k}) = 
-\frac{Z}{8\pi s\sqrt{v^2-s^2}}
\int_{-\omega_+}^\infty\frac{n(\epsilon')d\epsilon'}
{\sqrt{(\epsilon'+\omega_+)(\epsilon'+\omega_-)}}.
\end{equation}
The self-energy does not vary with $\omega$ if $n(\epsilon')$ is replaced 
with the step-function.  This channel of decay is characterized by the rate
\begin{equation}
-{\rm Im}\;\Sigma_{\rm sp}(\omega,{\bf k}) = 
\frac{3a\overline{|\Delta|^2}}{\epsilon_F},
\end{equation}
where $\sinh{a}=\sqrt{1/2}-1/2$.  
In $d=2+\varepsilon$ dimensions, $\overline{|\Delta|^2}\ll\epsilon_F^2$ 
and this additional linewidth is small in comparison with the broadening 
caused by induced fluctuations, of order $\sqrt{\overline{|\Delta|^2}}$.  
Therefore, the effect of fast spontaneous fluctuations can be neglected 
in $d=2+\varepsilon$ dimensions.


\vskip 1cm

\vskip 1cm

\vskip 1cm

\end{document}